# Blockchain security for ransomware detection

**Elodie Ngoie Mutombo**
u22608754@tuks.co.za
Department of Computer Science, University of Pretoria

**Mike Wa Nkongolo**
mike.wankongolo@up.ac.za
Department of Informatics, University of Pretoria

**Abstract**
Blockchain networks are critical for safeguarding digital transactions and assets, but they are increasingly targeted by ransomware attacks exploiting zero-day vulnerabilities. Traditional detection techniques struggle due to the complexity of these exploits and the lack of comprehensive datasets. The UGRansome dataset addresses this gap by offering detailed features for analysing ransomware and zero-day attacks, including timestamps, attack types, protocols, network flows, and financial impacts in bitcoins (BTC). This study uses the Lazy Predict library to automate machine learning (ML) on the UGRansome dataset. The study aims to enhance blockchain security through ransomware detection based on zero-day exploit recognition using the UGRansome dataset. Lazy Predict streamlines different ML model comparisons and identifies effective algorithms for threat detection. Key features such as timestamps, protocols, and financial data are used to predict anomalies as zero-day threats and to classify known signatures as ransomware. Results demonstrate that ML can significantly improve cybersecurity in blockchain environments. The DecisionTreeClassifier and ExtraTreeClassifier, with their high performance and low training times, are ideal candidates for deployment in real-time threat detection systems.

## 1. Introduction

Blockchain networks have revolutionised digital transactions by providing a decentralised, immutable ledger that enhances security and transparency [1]. They are widely used in various applications, including cryptocurrencies, supply chain management, and smart contracts [1], [2]. Despite their advanced security features, blockchain networks are not immune to attacks [3]. One significant vulnerability is the risk of ransomware attacks [4], [5], [6].

These attacks involve malicious actors encrypting data and demanding a ransom for decryption keys [7]. For instance, the WannaCry ransomware attack in 2017 exploited vulnerabilities in unpatched Windows systems, encrypting files and demanding Bitcoin payments [8]. While this specific attack was not directly on blockchain networks, it highlights how ransomware can disrupt digital operations, including those involving blockchain.

Another critical threat is zero-day exploits, which involve attacks leveraging previously unknown vulnerabilities [9]. For example, if a blockchain-based platform has a flaw in its smart contract code that has not yet been discovered, attackers can exploit this vulnerability to manipulate transactions or steal assets.

The DAO hack in 2016 is a notable example where an exploit in the code of a decentralised autonomous organisation led to the theft of approximately $60 million in Ethereum [10]. Denial of Service (DoS) attacks also pose a threat, where attackers overwhelm the network with excessive requests, causing disruptions [11]. An example is the Ethereum network congestion caused by spam transactions in 2016, which slowed down the network and impacted transactions [12]. While blockchain technology offers robust mechanisms to secure digital transactions, understanding and addressing these vulnerabilities is crucial for maintaining the integrity and reliability of blockchain networks [10], [11], [2]. As blockchain technology evolves, so too must the strategies for defending against these sophisticated and emerging threats [13].

## 1.1. Challenges in traditional detection techniques

Traditional detection techniques often struggle with blockchain security due to the inherent complexity of zero-day exploits and the scarcity of comprehensive datasets [14]. Zero-day exploits, which target previously unknown vulnerabilities, present a significant challenge because conventional security solutions rely on known threat signatures and patterns [15]. For instance, traditional antivirus software and intrusion detection systems (IDS) are designed to detect known threats by comparing observed behaviours against a database of signatures. However, zero-day exploits bypass these mechanisms because they exploit vulnerabilities that are not yet documented or recognised.

In the context of blockchain networks, the problem is compounded by the decentralised nature of the technology and the diverse types of attacks it faces [1], [2]. For example, an attacker might exploit a flaw in a smart contract to execute unauthorised transactions or manipulate contract logic. Since this flaw is unknown until discovered, traditional detection methods cannot preemptively identify or mitigate the threat.

Moreover, the lack of comprehensive datasets exacerbates the issue [16], [17]. Effective threat detection and response require extensive and well-curated datasets that capture a wide range of attack scenarios and behaviours [6], [18]. Nevertheless, in the blockchain security landscape, datasets that include detailed records of various attack vectors and anomalies are often sparse or non-existent [6], [16], [17], [18]. This limitation hampers the ability to train and validate detection models effectively.

## 1.2 Dataset features and their role in enhancing blockchain security

A well-structured dataset with comprehensive features can significantly enhance blockchain security by providing valuable insights into attack patterns and vulnerabilities [6], [18]. Table 1 outlines the features that can be utilised for detecting ransomware and zero-day exploits in blockchain environments. Machine learning (ML) models can be trained using these features to detect anomalies, classify known threats, and identify new attack signatures [19]. For example, using a dataset with detailed features like those mentioned in Table 1, a ML model could learn to recognise specific patterns associated with ransomware or zero-day exploits and improve its ability to predict and mitigate emerging threats [20].

While traditional detection techniques face significant challenges in the context of blockchain security, comprehensive datasets with rich features offer a pathway to more effective threat detection and response [21].

These datasets play a crucial role in strengthening blockchain security by detecting and mitigating threats such as ransomware and zero-day attacks [20], [21]. Moreover, existing datasets are often employed to train ML models and develop detection algorithms, but these datasets come with limitations that hinder their effectiveness in addressing emerging threats (see Table 2).

| Features | Description | Advantage |
| --- | --- | --- |
| **Timestamps** | Track the timing of transactions and attacks, helping to identify unusual patterns or correlations that may indicate an ongoing attack. | Anomalies detected in transaction timings can signal potential zero-day exploits or unauthorised activities. |
| **Protocols** | Detail the network communication methods used in blockchain transactions. | By analysing protocol data, NIDS can identify deviations from standard behaviours that might suggest an exploit or vulnerability. |
| **Financial data** | Transaction amounts and cryptocurrency values provide insights into the impact and scale of attacks. | Unusual spikes in Bitcoin transactions may indicate ransomware activity, allowing for timely intervention. |
| **Attack types and flags** | Categorise the nature of threats and connection statuses. | This classification helps in recognising known attack patterns and differentiating between legitimate and malicious activities. |
| **Network flows and IP addresses** | Offer visibility into the data transfer patterns and the origins of network requests. | Anomalies in network flows or suspicious IP addresses can highlight potential attacks or compromised nodes. |

**Table 1.** Dataset features in the blockchain security landscape

The UGRansome dataset offers a valuable advancement in addressing these limitations [22]. This dataset, created by Nkongolo et al. in 2021 [23], has been utilised in subsequent studies by Tokmak [9] and Alhashmi et al. [22]. UGRansome provides a comprehensive set of features specifically designed for analysing ransomware and zero-day attacks (Figure 1), filling a critical gap in blockchain security datasets. With its detailed features shown in Figure 1, it improves the detection and classification of ransomware and zero-day attacks compared to existing datasets (Table 2) and offers a more targeted approach to analysing blockchain-specific threats [24]. This dataset is bridging the gap left by traditional datasets by including salient and novel features relevant for network security in blockchain environments [7].

**Figure 1.** The UGRansome dataset

| Dataset | Advantage | Limitation |
|---|---|---|
| **VirusShare and MalwareBazaar** | Popular for malware analysis and ransomware detection. They include samples of known malware, which can be used to train models for detecting previously identified threats. | Since these datasets primarily focus on known malware, they lack information on new, previously unknown exploits that can compromise blockchain security. |
| **CICIDS and CTU-13** | Used for NIDS and include various types of attacks, such as DDoS and port scanning. | Less effective in addressing blockchain-specific threats. |
| **BTC-ETH Blockchain Data** | Some datasets provide transaction data from Bitcoin and Ethereum blockchains. They include details on transaction amounts, timestamps, and addresses. | Lack detailed annotations related to specific types of attacks, such as ransomware or zero-day exploits. Their granularity and scope are insufficient for comprehensive threat detection. |

**Table 2.** Existing datasets limitations in blockchain security

## 2. Methodology

In this study, Lazy Predict was used to simplify the process of model selection by automating four key steps in the ML workflow [25]:

1. *Model training and evaluation*

Lazy Predict uses various algorithms to fit ML models to the UGRansome dataset and evaluate their performance. The specific training objective function can mathematically represent each algorithm. For instance:

- **Linear regression**: The objective was to minimise the Mean Squared Error (MSE).

$$MSE = \frac{1}{n}\sum_{i=1}^{n}(y_i - \hat{y}_i)^2$$

Where y is the actual and predicted value, and n is the number of observations.

- **Decision trees**: Use criteria like Gini impurity or entropy to split nodes.

$$Gini(t) = 1 - \sum_{i=1}^{k} p_i^2$$

Where p is the probability of an instance being classified into class i at node t, and k is the number of classes.

- **Support vector machines (SVM)**: Aim to maximise the margin between classes.

$$Margin = \frac{2}{||w||}$$

Where w is the weight vector defining the separating hyperplane.

*2. Cross-validation*

Lazy Predict often uses cross-validation to assess ML model performance. Cross-validation involved partitioning the dataset into k subsets (folds) and performing training and evaluation k times, each time using a different fold as the validation set [25]. The cross-validation error is computed as follows:

$$CV\ Error = \frac{1}{k}\sum_{j=1}^{k} Error_j$$

where Error is the error of the ML model on the j-th fold.

*3. Performance metrics*

The study utilised accuracy, balanced accuracy, F1-score, and Time Taken to evaluate the Lazy Predict models.

*4. Model selection*

Lazy Predict automated the comparison of various models by computing these metrics and presenting them in a tabular format. The automation process involves:

- **Instantiation of ML models**: Each ML model is created with default parameters.
- **Training**: ML models are trained on the training data.
- **Evaluation**: ML models are evaluated using cross-validation or a validation set.
- **Comparison**: ML metrics are computed and compared to select the best-performing model.

When applied to the UGRansome dataset, Lazy Predict automates the application of evaluation metrics across multiple algorithms to significantly reduce the manual effort required for initial model experimentation [25]. The efficiency and effectiveness of each ML model are assessed based on the above metrics, providing a streamlined approach to model selection and optimisation.

## 3. Ransomware Analysis using Lazy Predict

Figure 2 depicts the 17 ransomware families included in the UGRansome dataset [4].

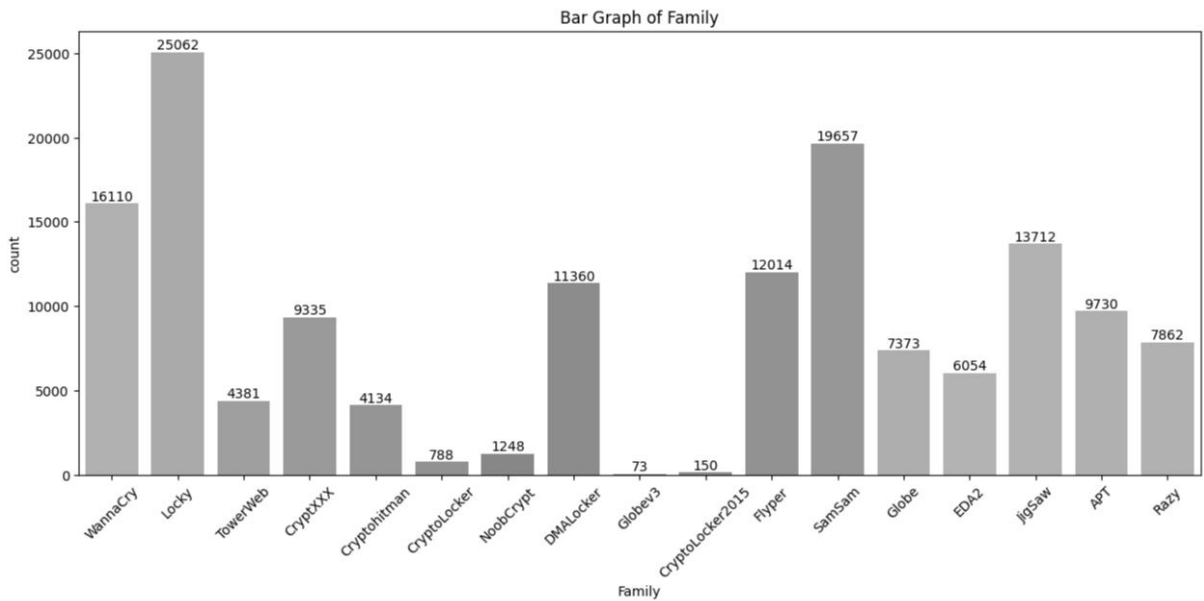

**Figure 2.** Ransomware family in the UGRansome dataset

The correlation of features is depicted in Figure 3. Understanding the significance of higher correlations of specific features in the context of ransomware detection is crucial for enhancing blockchain security. This study focuses on features such as network bytes, Bitcoin (BTC) transactions, threats, addresses, USD transactions, and ports to develop a robust ML model for zero-day exploit recognition (Figure 3). An in-depth analysis of higher correlations of these features is illustrated in Figure 3. Higher correlations of these features with ransomware detection underscore their importance in recognising zero-day exploits.

Figure 3 portrays these key indicators for network bytes, BTC transactions, threats, addresses, USD transactions, and ports (Figure 3). A high correlation between bitcoins (BTC) transactions and ransomware incidents suggests that transactions in BTC are a key marker of ransomware activity.

*Implication*: Since ransomware often demands payment in BTC due to its pseudonymous nature, tracking BTC transactions can be an effective way to identify and trace ransomware-related financial activities. This can provide critical insights into the financial trails of ransomware operations. In addition, a high correlation with USD transactions suggests that monetary transactions in USD are linked to ransomware activities. Ransomware may involve financial transactions in USD, either directly or indirectly.

Monitoring these transactions can help in identifying unusual patterns that may indicate ransomware activity, thus providing another layer of financial analysis for threat detection.

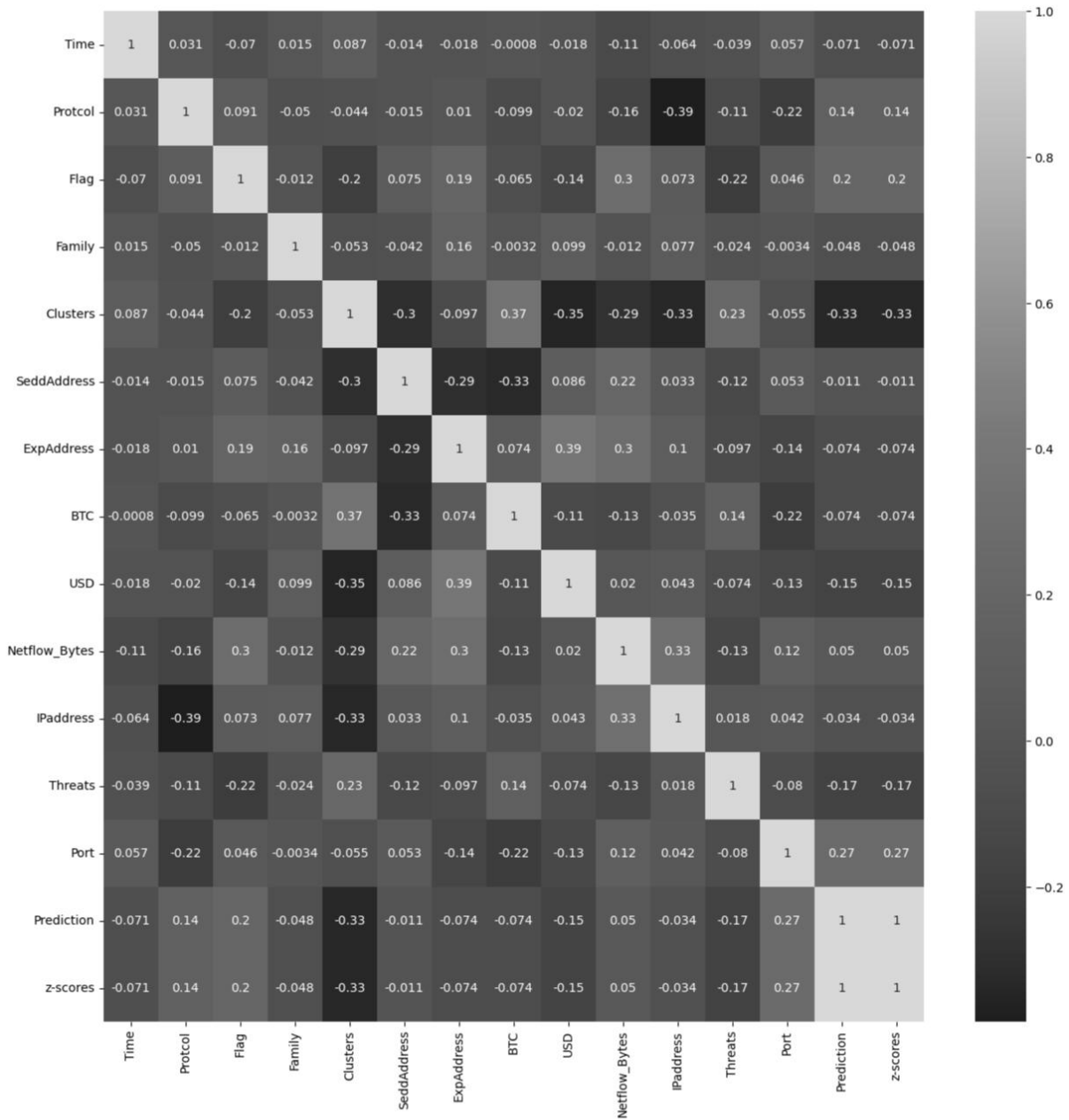

**Figure 3.** Feature correlation using the UGRansome dataset

## 4. Discussion on Lazy Predict results and ROC analysis

The results obtained from Lazy Predict provide a comprehensive overview of the performance of various ML models on the UGRansome dataset (Figure 4). Each model was evaluated based on accuracy, balanced accuracy, F1-score, and time taken, with a notable absence of the Receiver Operating Characteristic (ROC) metric (Figure 4). This absence can be attributed to the configuration settings within the Lazy Predict library, which might not have included the calculation or plotting of the ROC curve by default (Figure 4).

| Model | Accuracy | Balanced Accuracy | ROC AUC | F1 Score | Time Taken |
|---|---|---|---|---|---|
| LGBMClassifier | 1.00 | 0.99 | None | 1.00 | 4.22 |
| XGBClassifier | 0.99 | 0.99 | None | 0.99 | 5.45 |
| BaggingClassifier | 0.99 | 0.99 | None | 0.99 | 1.91 |
| DecisionTreeClassifier | 0.99 | 0.99 | None | 0.99 | 0.30 |
| RandomForestClassifier | 0.99 | 0.99 | None | 0.99 | 6.83 |
| ExtraTreesClassifier | 0.99 | 0.99 | None | 0.99 | 5.77 |
| ExtraTreeClassifier | 0.99 | 0.98 | None | 0.99 | 0.12 |
| KNeighborsClassifier | 0.98 | 0.98 | None | 0.98 | 2.72 |
| SVC | 0.97 | 0.96 | None | 0.97 | 44.99 |
| NuSVC | 0.89 | 0.87 | None | 0.89 | 551.55 |
| QuadraticDiscriminantAnalysis | 0.87 | 0.86 | None | 0.87 | 0.35 |
| AdaBoostClassifier | 0.86 | 0.85 | None | 0.86 | 3.83 |
| SGDClassifier | 0.82 | 0.79 | None | 0.81 | 1.26 |
| CalibratedClassifierCV | 0.81 | 0.77 | None | 0.80 | 174.08 |
| RidgeClassifier | 0.81 | 0.77 | None | 0.80 | 0.23 |
| RidgeClassifierCV | 0.81 | 0.77 | None | 0.80 | 0.53 |
| LogisticRegression | 0.80 | 0.77 | None | 0.79 | 1.31 |
| GaussianNB | 0.79 | 0.76 | None | 0.79 | 0.09 |
| BernoulliNB | 0.78 | 0.76 | None | 0.78 | 0.27 |
| NearestCentroid | 0.75 | 0.74 | None | 0.76 | 0.19 |
| PassiveAggressiveClassifier | 0.72 | 0.67 | None | 0.70 | 0.43 |
| Perceptron | 0.71 | 0.67 | None | 0.69 | 0.49 |
| DummyClassifier | 0.39 | 0.33 | None | 0.22 | 0.06 |

**Figure 4.** Lazy Predict results

To address this, we have computed the ROC and plotted the ROC curve (Figure 5) by fitting the models separately using appropriate libraries such as scikit-learn. The ROC curve is a crucial tool for evaluating the performance of classification models, especially in distinguishing between classes. It provides a graphical representation of a model's sensitivity (True Positive Rate) versus Specificity (False Positive Rate) across different threshold settings. For a comprehensive analysis, the ROC AUC (Area Under the Curve) metric can potentially quantify the overall ability of the ML model to discriminate between positive and negative classes. Given its importance, manually calculating the ROC AUC for the best-performing models can offer deeper insights into their performance, particularly in the context of zero-day exploit recognition and ransomware detection in blockchain security.

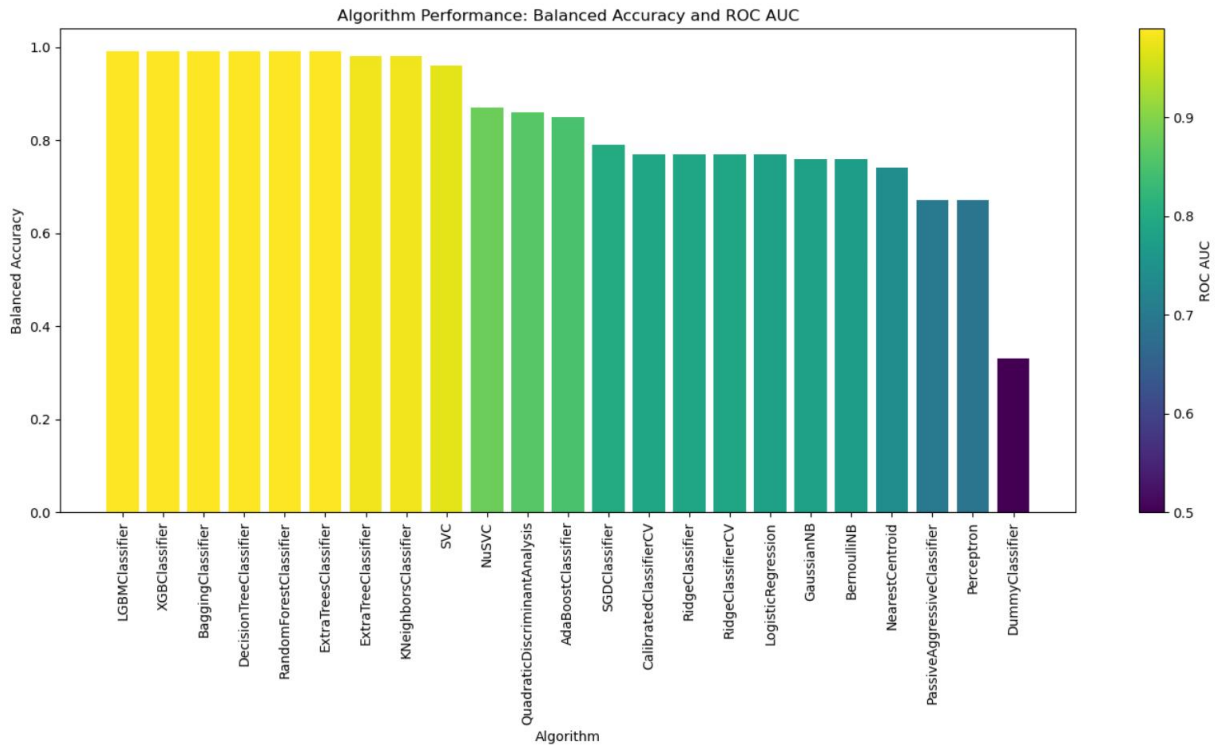

**Figure 5**. Balance accuracy and ROC AUC

Among the ML models tested, the LGBMClassifier stands out with its perfect accuracy and F1-score, making it highly effective for ransomware detection (Figure 4 and Figure 5). However, the time taken for model training is also a critical factor. The DecisionTreeClassifier, with an accuracy and F1-score of 0.99 and a minimal training time of 0.30 seconds, emerges as a highly efficient model (Figure 4 and Figure 5). Similarly, the ExtraTreeClassifier demonstrates comparable performance with an even shorter training time of 0.12 seconds (Figure 4 and Figure 5). These ML models are ensemble learning techniques, particularly valuable in real-time blockchain security applications where rapid detection and response to ransomware attacks are crucial [22].

The efficiency of these models in detecting ransomware is evident from the results presented in Figure 6. These ML models are a powerful way to find and stop threats in real time because they can accurately tell the difference between zero-day exploits (encoded as 0 using Python's label encoder), known ransomware (encoded as 1), and synthetic threats (encoded as 2). The models successfully predict 4,748 features as anomalies, representing zero-day attacks (Figure 6). This capability is crucial for blockchain security as it allows for the identification of previously unknown threats that traditional detection methods might miss. Additionally, the models classify 7,455 features as signatures, indicating known ransomware activities and facilitating prompt recognition and response (Figure 6). Furthermore, 6,761 features are identified as synthetic signatures, likely representing variations or simulations of known ransomware patterns (Figure 6). This comprehensive detection across various categories underscores the models' robustness and effectiveness in enhancing cybersecurity within blockchain networks.

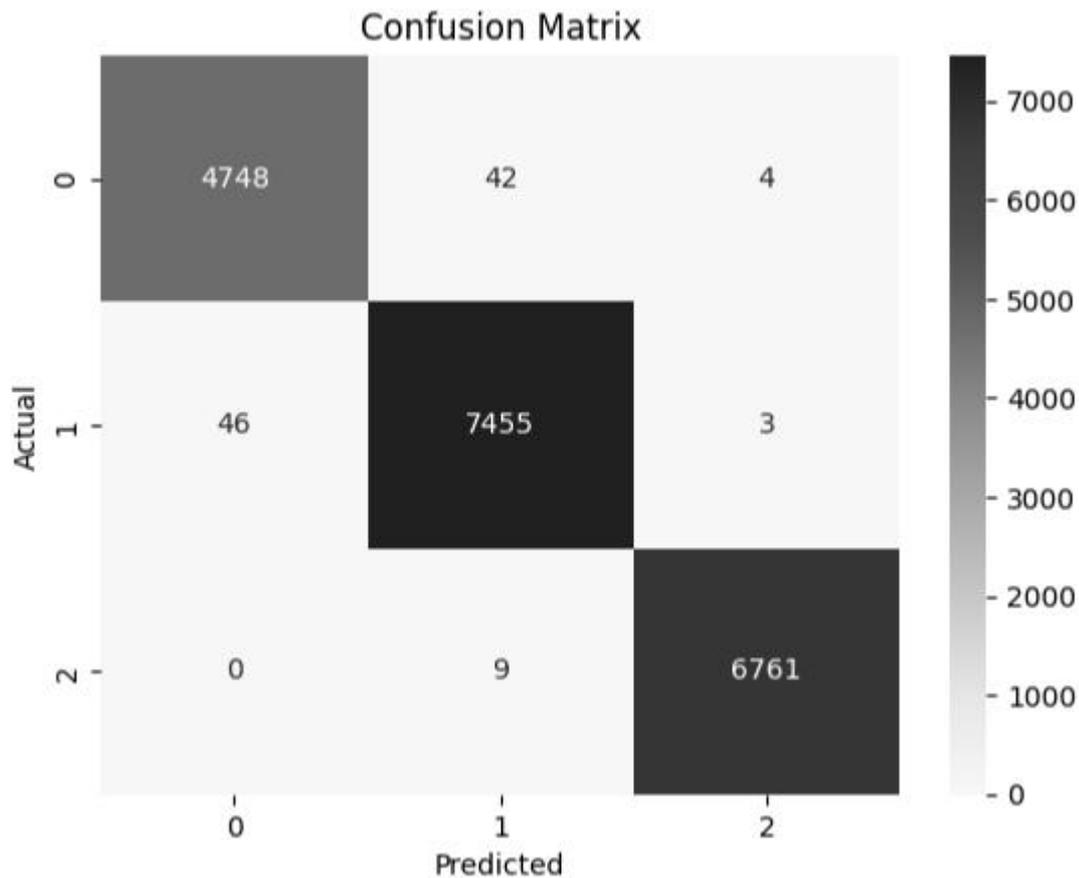

**Figure 6.** The ensemble learning (BaggingClassifier) confusion matrix

Features such as network bytes, BTC transactions, and protocol data, can be used by these models to identify anomalies indicative of zero-day exploits and ransomware activities. The quick training times of DecisionTreeClassifier and ExtraTreeClassifier ensure that the models can be frequently updated with new data to maintain their effectiveness against evolving threats (Figure 4).

On the other hand, models like the NuSVC, while accurate, have prohibitively long training times (551.55 seconds), making them less practical for real-time applications (Figure 4). Similarly, the SVC, despite its high accuracy, has a significant training time of 44.99 seconds, which may not be suitable for environments where rapid response is required (Figure 4).

In conclusion, the results from Lazy Predict highlight the importance of balancing accuracy and computational efficiency in selecting ML models for ransomware detection in blockchain security. Ensemble learning techniques, with their high performance and low training times, are ideal candidates for deployment in real-time threat detection systems [9], [22]. To further enhance this analysis, calculating and plotting the ROC AUC for these ML models will provide a more detailed understanding of their discriminative capabilities.

## 5. Conclusion

Blockchain networks are critical for safeguarding digital transactions and assets, but they are increasingly targeted by ransomware attacks exploiting zero-day vulnerabilities. Traditional detection techniques struggle due to the complexity of these exploits and the lack of comprehensive datasets. The UGRansome dataset addresses this gap by offering detailed features for analysing ransomware and zero-day attacks.

This study employs the Lazy Predict library to automate ML on the UGRansome dataset, aiming to enhance blockchain security through ransomware detection based on zero-day exploits recognition. Lazy Predict streamlines the comparison of different ML models and identifies effective algorithms for threat detection. Key features such as timestamps, protocols, and financial data are utilised to predict anomalies as zero-day threats and classify known signatures as ransomware.

Results demonstrate that ML models can significantly improve cybersecurity in blockchain environments. Among the models tested, the DecisionTreeClassifier and ExtraTreeClassifier, with their high performance and low training times, emerge as ideal candidates for deployment in real-time threat detection systems. These findings underscore the potential of advanced ML techniques in fortifying blockchain security and provide a robust framework for future research in anomaly detection and cybersecurity analytics.

## Dataset and code availability

Mike Nkongolo Wa Nkongolo (2023). UGRansome dataset. Kaggle. Available at: https://www.kaggle.com/dsv/7172543 [Accessed 23 July 2024]. DOI: 10.34740/KAGGLE/DSV/7172543.

## Acknowledgement

This work is part of a master's in computer science proposal at the University of Pretoria.